\documentclass{svjour3}                     
\smartqed  
\usepackage{graphicx}
\usepackage{mathptmx}      
\usepackage{natbib}
\usepackage{url}
\usepackage{slatex} 
\usepackage{mmm}
\usepackage{mathpartir}
\usepackage{epsf}
  \let\cite=\citep

\usepackage[
  pdfauthor={Sam Tobin-Hochstadt and Matthias Felleisen},
  pdftitle={The Design and Implementation of Typed Scheme},
  pdfsubject={Computer Science}]
  {hyperref}

\setkeyword{require/typed pdefine: Listof require-typed-struct lib 
  planet Symbol Opaque require/opaque-type define-values values
  All cases define-type Any Refinement declare-refinement provide
  -> Number String Boolean : define: define-type-alias define-struct
  define-typed-struct module Rec Pair Integer define-struct:
  #
\setspecialsymbol{lambda}{\l} 
\setspecialsymbol{mu}{$\mu$} 
\setspecialsymbol{->}{$\to$}
\setspecialsymbol{lamU}{$\lambda_{U}$}
\setspecialsymbol{lamC}{$\lambda_{K}$}
\setspecialsymbol{lamV}{$\lambda_{V}$}
\setspecialsymbol{lam}{$\lambda$}
\setspecialsymbol{lamA}{$\lambda_A$}
\setspecialsymbol{Lam}{$\Lambda$}
\setspecialsymbol{UserVar}{$V_u$}
\setspecialsymbol{ContVar}{$V_k$}
\setspecialsymbol{AnyVar}{$V$}
\setspecialsymbol{U}{\ma{\bigcup}}
\setspecialsymbol{-U>}{\ma{\stackrel{\mbox{(\usym\ {\bf Number} {\bf Boolean})}}{\to}}}
\setspecialsymbol{bover}{\ma{\stackrel{b}{\to}}}
\setspecialsymbol{bot}{\ma{\bot}}
\setspecialsymbol{lang}{\ma{\#{\bf lang}}}
\setspecialsymbol{ELIDED}{\mbox{$\mbox{\tiny ~~{\it omitted\/}~~} \over ~$}}

\leftcodeskip=15pt

\journalname{Higher-Order and Symbolic Computation}

\begin{document}

\begin{schemeregion}

\input{sample.sty}

\title{The Design and Implementation of Typed Scheme: \\
From Scripts to
  Programs\footnote{This is a revised and extended version of a paper
    presented at the 35th ACM SIGPLAN-SIGACT Symposium on Principles
    of Programming Languages, 2008.  The authors were supported by
    the NSF, the USAF, and the Mozilla Foundation.}}

\titlerunning{The Design and Implementation of Typed Scheme}        

\author{Sam Tobin-Hochstadt         \and
        Matthias Felleisen 
}


\institute{S. Tobin-Hochstadt \at
              Northeastern University \\
              \email{samth@ccs.neu.edu}
           \and
           M. Felleisen \at
           Northeastern University
}

\date{Received: date / Accepted: date}

\maketitle

\begin{abstract}
When scripts in untyped languages grow into large programs, maintaining
 them becomes difficult. A lack of explicit 
 type annotations in typical scripting languages forces programmers to must
 (re)discover critical pieces of design information every time they wish to
 change a program. This analysis step both slows down the maintenance
 process and may even introduce mistakes due to the violation of
 undiscovered invariants. 

 This paper presents Typed Scheme, an explicitly typed extension of
 PLT
 Scheme, an
 untyped scripting language. Its type system is based on the novel notion
 of \emph{occurrence typing}, which we formalize and mechanically prove
 sound. The implementation of Typed Scheme additionally borrows elements
 from a range of approaches, including recursive types, true unions and
 subtyping, plus polymorphism combined with a modicum of local
 inference.

 The formulation of occurrence typing naturally leads to a simple and
 expressive version of predicates to describe \emph{refinement types}. 
 A Typed Scheme program can use these refinement types to keep track
 of arbitrary classes of values via the type system.  
Further, we show how the Typed Scheme type system, in conjunction with
simple recursive types, is able to encode refinements of existing
datatypes, thus expressing \emph{both} proposed variations of
refinement types.   

\end{abstract}

\section{Type Refactoring: From Scripts to Programs}


Recently, under the heading of ``scripting languages'', a variety of
new languages have become popular, and even pervasive, in web- and
systems-related fields.  Due to their popularity, programmers often
create scripts that then grow into large applications.

 Most scripting languages are untyped and provide
 primitives with flexible semantics
 to make programs concise.  Many programmers find these attributes
 appealing and use scripting languages for these reasons. 
 Programmers are also beginning to notice, however, that untyped scripts are
 difficult to maintain over the long run. The lack of types means a
 loss of design information that programmers must recover every time
 they wish to change existing code. Both the Perl community~\citep{tang} and the
 JavaScript community~\citep{ecma} are implicitly acknowledging this 
 problem by considering
 the addition of Common Lisp-style~\citep{cl} typing constructs to the
 upcoming releases of their respective languages.  Additionally,
 type system proposals have been made for Ruby~\cite{furr:ruby:sac}
 and for Dylan~\cite{dylan:diplom-thesis}

 In the meantime, industry faces the problem of porting existing
 application systems from untyped scripting languages to the typed
 world. In response, we have proposed a theoretical model
 for this conversion process and have shown that partial conversions can
 benefit from type-safety properties to the desired
 extent~\cite{thf:dls2006}. This problem has also sparked significant
 research interest in the evolution of scripts to programs~\cite{stop09}.
The key assumption behind our work is the
 existence of an explicitly typed version of the scripting language, with
 the same semantics as the original language, so that values can freely
 flow back and forth between typed and untyped modules.  In other words, we
 imagine that programmers can simply add type annotations to a module
 and thus introduce a certain amount of type-safety into the
 program.

 At first glance, our assumption of such a typed sister language
 may seem unrealistic.  Programmers
 in untyped languages often loosely mix and match reasoning from
 various type disciplines when they write scripts. Worse, an
 inspection of code suggests they also include flow-oriented
 reasoning, distinguishing types for variables 
 depending on prior operations. In short, untyped scripting
 languages permit programs that appear difficult to type-check with
 existing type systems.

To demonstrate the feasibility of our approach, we have designed and
implemented Typed Scheme, an explicitly typed version of PLT
Scheme. We have chosen PLT Scheme for two reasons. On one hand, PLT
Scheme is used as a scripting language by a large number of users. It
also comes with a large body of code, with contributions ranging from
scripts to libraries to large operating-system like programs.  On the
other hand, the language comes with macros, a powerful extension
mechanism~\cite{f:modules}. Macros place a significant constraint on
the design and implementation of Typed Scheme, since supporting macros
requires typechecking a language with a user-defined set of syntactic
forms.  We  are able to overcome this difficulty by integrating the
type checker with the macro
expander.  Indeed, this approach ends up greatly facilitating the
integration  of typed and untyped modules.
As envisioned \cite{thf:dls2006}, this integration
makes it mostly straightforward to turn portions of a multi-module program into a
partially typed yet still executable program.

Developing Typed Scheme requires not just integration with the
underlying PLT Scheme system, but also a type system that works well
with the idioms used by PLT Scheme programmers when developing
scripts.  It would be an undue burden if the programmer needed to
rewrite idiomatic PLT Scheme code to make it typeable in Typed Scheme.
For this purpose, we have developed a novel type system, combining the
idea of \emph{occurrence} typing with subtyping, recursive types,
polymorphism and a modicum of inference.  

The design of Typed Scheme and its type system also allows for simple
additions of sophisticated type system features.  In particular,
the treatment of predicates in Typed Scheme lends itself naturally to
treating predicates such as \scheme|even?| as defining refinements of
existing types, such a integers.  This allows for a lightweight form
of refinement types, without any need for implication or inclusion
checking.  

 We first present a formal model of
the key aspects of occurrence typing and prove it to be type-sound. We
then describe how refinement types can be added to this system, and
how they can be used effectively in Typed Scheme.  Later we describe
 how to scale this calculus into a full-fledged, typed version of PLT
 Scheme and how to implement it. Finally, we give an account of our preliminary
 experience,
 adding types to thousands of lines of untyped Scheme code. Our
 experiments seem promising and suggest that converting untyped scripts
 into well-typed programs is feasible.


\section{Overview of Typed Scheme}
\label{s:overview}

The goal of the Typed Scheme project is to develop an explicit type system
that easily accommodates a conventional Scheme programming style. Ideally,
programming in Typed Scheme should feel like programming in PLT Scheme,
except for typed function and structure signatures plus  type
definitions. Few other changes should be required when going from a Scheme
program to a Typed Scheme program.   Furthermore, the addition of types should
require a relatively small effort, compared to the original program.
This requires that macros, both those used and defined in the typed
program, must be supported as much as possible. 

Supporting this style of programming demands a significant rethinking
of type systems.  Scheme programmers reason
about their programs, but not with any conventional type system
in mind. They superimpose on their untyped syntax whatever type (or
analysis) discipline is convenient.  No existing type system could cover
all of these varieties of reasoning.  

\subsection{Occurrence Typing}

Consider the following function
definition:\footnote{Standards-conforming Scheme implementations
  provide a complex number datatype directly.  This example serves only expository purposes.}
\begin{schemedisplay}
;; data definition: a Complex is either 
;; - a Number or 
;; - (cons Number Number)

;; Complex $\rightarrow$ Number
(define (creal x)
  (cond [(number? x) x]
	[else (car x)]))
\end{schemedisplay}
As the informal data definition states, complex numbers are
represented as either a single number, or a
pair of numbers (cons).

The definition illustrates several key elements of the way that Scheme
programmers reason about their programs: ad-hoc type specifications,
true union types, and predicates for type testing.  No
datatype specification is needed to introduce a sum type on which the
function operates. Instead there is just an ``informal'' data
definition and contract~\cite{fffk:htdp}, which gives a name to a
set of pre-existing data, without introducing
new constructors. Further, the function does not
use pattern matching to dispatch on the union type. Instead, it uses a
predicate that distinguishes the two cases: the first \scheme|cond|
clause, which treats \scheme{x} as a number and the second one,
which treats it as a pair.

Here is the corresponding Typed Scheme code:
\begin{schemedisplay}
(define-type-alias Cplx (U Number (cons Number Number)))

(define: (creal [x : Cplx]) : Number
  (cond [(number? x) x]
	[else (car x)]))
\end{schemedisplay}
This version explicates both aspects of our informal reasoning.
The type \scheme|Cplx| is an abbreviation for the true union intended by
the programmer; naturally, it is unnecessary to introduce type
abbreviations like this one. Furthermore, the body of \scheme{creal} is not
modified at all; Typed Scheme type-checks each branch of the conditional
appropriately. In short, only minimal type annotations are required to
obtain a typed version of the original code, in which the informal,
unchecked comments become statically-checked design elements.

Our design also accommodates more complex reasoning about the flow of
values in Scheme programs.  

\begin{schemedisplay}
(foldl scene+rectangle empty-scene (filter rectangle? list-of-shapes))
\end{schemedisplay}
This code selects all the \scheme|rectangle|s from a list of shapes,
and then adds them one by one to an initially-empty scene, perhaps
in preparation for rendering to the screen.  Even though the initial
\scheme|list-of-shapes| may contain shapes that are not
\scheme|rectangle|s, those are removed by the \scheme|filter|
function.  The resulting list contains only \scheme|rectangle|s, and
is an appropriate argument to \scheme|scene+rectangle|.  No additional
coercions are needed.

This example demonstrates a different mode of reasoning than the first;
 here, the Scheme programmer uses polymorphism and the
argument-dependent invariants of \scheme|filter|  to ensure correctness.  

No changes to this code are required for it to typecheck in Typed
Scheme. The type system is able to accommodate both  modes of
reasoning the programmer uses with polymorphic functions and
occurrence typing. In contrast, a more conventional type system such
as SML~\cite{sml:bible} would require
the use of an intermediate data type, such as an option type, to
ensure conformance.










\subsection{Refinement Types}

Refinement types, introduced originally by \citet{fp:refinement}, are
types which describe subsets of conventional types.  For example, the
type of even integers is a refinement of the type of integers.  Many
different  systems have proposed distinct ways of specifying these
subsets~\cite{liquid-types,wadler-findler}.  In Typed Scheme, we describe a set of values with a simple
Scheme predicate.  

The fundamental idea is that a boolean-valued function, such as
\scheme|even?|, can be treated as defining a type, which is a subtype
of the input type of \scheme|even?|.  This type has no constructors,
but it is trivial to determine if a value is a member by using the
predicate \scheme|even?|.  For example, this function produces solely
even numbers:\footnote{In the subsequent formal development, we
  require a slightly more verbose syntax for refinement types.}

\begin{schemedisplay}
(: just-even (Number -> (Refinement even?)))
(define (just-even n)
  (if (even? n) n (error 'not-even)))
\end{schemedisplay}

This technique harnesses occurrence typing to work with arbitrary
predicates, and not just those that correspond to Scheme data types.

\subsection{Other Type System Features}

In order to support Scheme idioms and programming styles, Typed Scheme
supports a number of type system features that have been studied
previously, but are rarely found in a single, full-fledged implementation.
Specifically, Typed Scheme supports true union
types~\cite{pierce-union}, as seen above.  It also
provides first-class polymorphic functions, known as impredicative
polymorphism, a feature of the
Glasgow Haskell Compiler~\cite{rank-n}.  In addition, Typed Scheme
allows programmers to explicitly specify recursive types, as well as
constructors and accessors that manage the recursive types
automatically.  Finally, Typed Scheme provides a rich set of base
types to match those of PLT Scheme.






\subsection{S-expressions}

One of the primary Scheme data structures is the
S-expression.  We have already seen an example of this in the
preceding section, where we used pairs of numbers to represent complex
numbers.  Other uses of S-expressions abound in real Scheme code,
including using lists as tuples, records, trees, etc.  Typed Scheme
handles these features by representing lists explicitly as
sequences of \scheme|cons| cells.  Therefore, we can give an
S-expression as precise a type as desired.  For example, the
expression \scheme|(list 1 2 3)| is given the type 
\scheme|(cons Number (cons Number (cons Number '())))|, which is a
subtype of \scheme|(Listof Number)|.

Lists, of course, are recursive structures, and we exploit Typed
Scheme's support for explicit recursive types to make \scheme|Listof|
a simple type definition over \scheme|cons|.  Thus, the subtyping
relationship for fixed-length lists is simply a consequence of the
more general rules for recursive types.

Sometimes, however, Scheme programmers rely on invariants too subtle
to be captured in our type system.  For example, S-expressions are
often used to represent XML data, without first imposing any structure
on that data.  In these cases, Typed Scheme allows programmers to
leave the code dealing with XML in the untyped world, communicating
with the typed portions of the program just as other untyped code does.


\subsection{Other Important Scheme Features}

Scheme programmers also use numerous programming-language features
that are not present in typical typed languages.  Examples
of these include the \scheme|apply| function, which applies a function
to a heterogeneous list of arguments; the multiple value return
mechanism in Scheme; the use of arbitrary non-false values in
conditionals; the use of variable-arity and multiple-arity
functions; and many others.  Some variable-arity functions, such as
\scheme|map| and \scheme|foldl|, require special care in the type
system~\cite{sthf:variable-arity}. All of these features are
widely used in existing PLT Scheme programs, and supported by Typed Scheme.


\subsection{Macros}

Handling macros well is key for any system that claims to allow
typical Scheme practice.  This involves handling macros defined in
libraries or by the base language as well as macros defined in modules that are converted to
Typed Scheme.  Further, since macros can be imported from arbitrary
libraries, we cannot specify the typing rules for all macros ahead of
time.  Therefore, we must expand macros before typechecking.  This
allows us to handle almost all simple macros, and many existing
complex macros
 without change,
i.e., those for which we can infer the types of the generated variables.
Further, macros defined in typed code require no changes.
Unfortunately, this approach does not scale to the largest and most
complex macros, such as those defining a class
system~\cite{fff:classes}, which rely on and enforce their own
invariants that are not understood by the type system.  Handling such
macros remains future work.

\section{Two  Examples of Refinements}

\label{sec:example}

To demonstrate the utility of refinement types as provided by Typed
Scheme, as well as the other features of the language,
 we present two extended examples.  The first tackles the problem of form
validation, demonstrating the use of predicate-based refinements.  The
second encodes the syntax of the continuation-passing-style
$\lambda$-calculus in the type system.

\subsection{Form Validation}

  One important problem in form validation is avoiding SQL
injection attacks, where a piece of user input is allowed to contain
an SQL statement and passed directly to the database.  A simple 
example is the query

\begin{schemedisplay}
(string-append "SELECT * FROM users WHERE name = '" user-name "';")
\end{schemedisplay}

If \scheme|user-name| is taken directly from user input, then it might
contain the string \scheme|"a' or 't'='t"|, resulting in an query that
returns the entire contents of the \verb|users| table.  More damaging
queries can be constructed, with data loss a significant
possibility~\cite{xkcd}.  

One common solution for avoiding this problem is sanitizing user input
with escape characters.  Unfortunately, sanitized input, like
unsanitized input, is simply a string.  Therefore, we use
refinement types to statically verify that only validated input
 is passed through to the database.  This requires two key
pieces: the predicate, and the final consumer.

The predicate is a Typed Scheme function that determines if a
string is acceptable as input to the database:

\begin{schemedisplay}
(: sql-safe? (String -> Boolean))
(define (sql-safe? s) ELIDED)
\end{schemedisplay}  

No special type system machinery is required to write and use such a
predicate.  One more step is needed, however, to turn this predicate
into a refinement type:

\begin{schemedisplay}
(declare-refinement sql-safe?)
\end{schemedisplay}

\noindent
This declaration changes the type of \scheme|sql-safe?| to be
a predicate for \scheme|(Refinement sql-safe? String)|.\footnote{It is
 similar to the function \scheme|sql-safe?| being
in the environment $\Delta$ in the formalization of refinement types,
see section~\ref{sec:refinement}.}

With this refinement type, we can specify the desired type of our
query function:

\begin{schemedisplay}
(: query ((Refinement sql-safe? String) -> (Listof Result)))
(define (query user-name)
  (run-query
   (string-append "SELECT * FROM users WHERE name = '" user-name "';")))
\end{schemedisplay}

Since \scheme|(Refinement sql-safe? String)| is a subtype of
\scheme|String|, \scheme|user-name|  can be used directly as an
argument to \scheme|string-append|.  

We can also write a \scheme|sanitize| function that performs the
necessary escaping, and use the \scheme|sql-safe?| function and
refinement types for static and dynamic verification:

\begin{schemedisplay}
(: sanitize (String -> (Refinement sql-safe? String)))
(define (sanitize s)
  (define s* (string-map escape-char s))
  (if (sql-safe? s) s (error "escape failed")))
\end{schemedisplay}
\noindent
The only function that is added to the trusted computing
base is the definition of \scheme|sql-safe?|, which can be provided by
the database vendor.  Everything else can be entirely user-written.  

\paragraph{Alternative Solutions}

Another solution to this problem, common in other languages, would have \scheme|sanitize| be
defined in a different module, with \scheme|SQLSafeString| as an
opaque exported type.  Unfortunately, this requires using an accessor
whenever a \scheme|SQLSafeString| is used in a context that expects a
string (such as \scheme|string-append|).  The use of refinement types
avoids both the dynamic cost of wrapping in a new type, as well as the
programmer burden of managing these wrappers and their corresponding
accessors.

\subsection{Restricted Grammars}

Given a  recursive data type, it is common to
describe subsets of such data that are valid in a particular context.
Non-empty lists are a paradigmatic case, and are the original
motivating example for \citet{fp:refinement} in their work on
refinement types.   

In Typed Scheme, the type for a list of \scheme|Integer|s would be 
\begin{schemedisplay}
(define-type IntList (Rec L (U '() (Pair Integer L))))
\end{schemedisplay}
where \scheme|Rec| is the constructor for recursive types. 
Non-empty lists are just a single unfolding of this type, without the
initial \scheme|'()| case:
\begin{schemedisplay}
(define-type NonEmpty (Pair Integer (Rec L (U '() (Pair Integer L))))
\end{schemedisplay}
Of course, \scheme|NonEmpty| is a subtype of \scheme{IntList}.  

Using this technique, we can encode other interesting examples of
refinement types.  To demonstrate its expressiveness, we show how to
encode the \emph{partitioned CPS} of \citet[Definition 8]{sf:cps}.  We begin with
encodings of variables, which distinguishes variables ranging over
user values (\scheme|UserVar|) from variables ranging over
continuations (\scheme|ContVar|):

\begin{schemedisplay}
(define-type UserVar Symbol)
(define-struct: ContVar ([v : Symbol]))
(define-type AnyVar (U UserVar ContVar))
\end{schemedisplay}

The \scheme|lam| and \scheme|app| constructors are both parameterized
over their two field types.  A \scheme|lam| contains a variable and a
body, while an \scheme|app| has an operator and an operand:
\begin{schemedisplay}
(define-struct: (V A) lam ([x : V] [b : A]))
(define-struct: (A B) app ([rator : A] [rand : B]))
\end{schemedisplay}

\begin{schemedisplay}
(define-type (lamU A) (lam UserVar A))
(define-type (lamC A) (lam ContVar A))
(define-type (lamA A) (lam AnyVar A))
\end{schemedisplay}
\scheme|lamU| and \scheme|lamC| are abbreviations for
user-level and transformation-introduced abstractions.  \scheme|lamA|
allows any kind of variable.

With these preliminaries in place, we can define $\lambda$-terms:
\begin{schemedisplay}
(define-type Lam (Rec L (U AnyVar (lamA L) (app L L))))
\end{schemedisplay}
A term is either a variable (\scheme|AnyVar|), an abstraction whose
body is a term, or an application of two terms.

We now transform the original definition of
partitioned CPS terms:

\[
\begin{altgrammar}
P & ::= & \comb{K}{W} \\
W & ::= & x \alt \lambda{k}.{K} \\
K & ::= & k \alt \comb{W}{K} \alt \lambda{k}.{K} \\
\end{altgrammar}
\]

\begin{schemeregion}
We can write these definitions directly as Typed Scheme types:%
\footnote{Unfortunately, Typed Scheme currently requires the
  equivalent, but more verbose definition of \scheme|P| and
  \scheme|K|, in which the other definitions are inlined:
\begin{schemedisplay}
(define-type P (app K (U UserVar (lamC K))))
(define-type K (Rec K (U ContVar (app (U UserVar (lamC K)) K) (lamU (app K (U UserVar (lamC K)))))))
\end{schemedisplay}
We are investigating how to admit the shorter syntax directly.}
\begin{schemedisplay}
(define-type P (app K W))
(define-type W (U UserVar (lamC K)))
(define-type K (U ContVar (app W K) (lamC K))
\end{schemedisplay}
All of the types are of course subtypes of \scheme|Lam|.  Thus,
compiler writers can
write typeful functions that manipulate CPS terms, while also
using more general functions that accept arbitrary terms (such as evaluators) on them.
\end{schemeregion}




\section{A Formal Model of Typed Scheme}
\label{s:formal}

Following precedent, we have distilled the novelty of our type system
into a typed lambda calculus, \lts. While Typed Scheme incorporates
many aspects of modern type systems, the calculus serves only as a
model of occurrence typing, the primary novel aspect of the type
system, in conjunction with true
union types and subtyping.  The latter directly interact with the
former; other features of the type system are mostly orthogonal to
occurrence typing.  This section first presents the syntax and dynamic
semantics of the calculus, followed by the typing rules and a
(mechanically verified) soundness result.

\def\lateff{latent predicate}
\def\lexeff{visible predicate}
\def\subpred{sub-predicate}

\subsection{Syntax and Operational Semantics}

\newtheorem{define}{Definition}

\newcommand\red{\to^*} 
\newcommand{\sembrack}[1]{[\![#1]\!]}

\newmeta\e{e}
\newmeta\y{y}
\newmeta\G{\Gamma}
\newmeta\E{E}
\newmeta\d{d}
\newmeta\b{b}
\newmeta\c{c}
\newmeta\L{L}
\newmeta\uu{u}
\newmeta\vv{v}
\newmeta\kk{k}
\newmeta\n{n}
\newmeta\ph{\phi}

\begin{figure*}[htbp]

\[
  \begin{altgrammar}
   \d{}, \e{}, \dots &::=& \x{}  \alt \comb{\e1}{\e2} \alt \cond{\e1}{\e2}{\e3} \alt \vv{}  &\mbox{Expressions} \\
   \vv{} &::=& \c{}  \alt \b{} \alt \n{} \alt \abs{\x{}}{\t{}}{\e{}}  & \mbox{Values} \\
   \c{} &::=& \addone \alt \numberp \alt \boolp \alt \procp \alt \nott & \mbox{Primitive Operations} 
   \vspace{1mm}
   \\
   \E{} &::=& [] \alt \comb{\E{}}{\e{}} \alt \comb{\vv{}}{\E{}} \alt \cond{\E{}}{\e2}{\e3} & \mbox{Evaluation Contexts}
   \vspace{1mm} \\
   
    \ph{} & ::=& \t{} \alt \noeffect & \mbox{Latent Predicates} \\
    \p{} &::=& \t{\x{}} \alt \x{} \alt \tt \alt \ff \alt \noeffect & \mbox{Visible Predicate} \\
   \s{},\t{} &::= & \top \alt \num \alt \ttt \alt \fft \alt {\proctype {\s{}} {\t{}} {\ph{}}} \alt (\usym\ \t{} \dots) 
   &\mbox{Types} 
   \end{altgrammar}
\]

\caption{Syntax} \label{f:syntax}
\end{figure*}

\begin{figure*}[htbp]

\[
\inferrule[T-Var]{}{\ghastyeff{\x{}}{\G{}(\x{})}{\x{}}}
\qquad
\inferrule[T-Num]{}{\ghastyeff{\n{}}{\num}{\tt}}
\qquad
\inferrule[T-Const]{}{\ghastyeff{\c{}}{\dt{\c{}}}{\tt}}
\]

\[
\inferrule[T-True]{}{\ghastyeff{\tt}{\bool}{\tt}}
\qquad
\inferrule[T-False]{}{\ghastyeff{\ff}{\bool}{\ff}}
\]

\begin{center}
\[
\inferrule[T-AbsPred]{\hastyeff{\G{},\hastype{\x{}}{\s{}}}{\e{}}{\t{}}{\s{\x{}}'}}
        {\ghastyeff{\abs{\x{}}{\s{}}{\e{}}}{\proctype {\s{}} {\t{}} {\s{}'}}{\tt}}
\qquad
\inferrule[T-Abs]{\hastyeff{\G{},\hastype{\x{}}{\s{}}}{\e{}}{\t{}}{\p{}}}
        {\ghastyeff{\abs{\x{}}{\s{}}{\e{}}}{\proctype {\s{}} {\t{}} {\noeffect}}{\tt}}
\]
\[
\inferrule[T-App]{\ghastyeff{\e1}{\t{}'}{\p{}}
\\\\
\ghastyeff{\e2}{\t{}}{\p{}'} \\\\ \subtype{\t{}}{\t0}
\\\\
{\subtype{\t{}'}{\proctype{\t0}{\t1}{\ph{}}}}}
{\ghastyeff{\comb{\e1}{\e2}} {\t1} {\noeffect}}
\qquad
\inferrule[T-AppPred]{\ghastyeff{\e1}{\t{}'}{\p{}}
\\\\
\ghastyeff{\e2}{\t{}}{\x{}} \\\\ \subtype{\t{}}{\t0}
\\\\
{\subtype{\t{}'}{\proctype{\t0}{\t1}{\s{}}}}}
{\ghastyeff{\comb{\e1}{\e2}} {\t1} {\s{\x{}}}}
\qquad
\inferrule[T-If]
        {\ghastyeff{\e1}{\t1}{\p1}
        \\\\
        \hastyeff{\G{} + \p1}{\e2}{\t2}{\p2}
        \\\\
        \hastyeff{\G{} - \p1}{\e3}{\t3}{\p3}
        \\\\
        \subtype{\t2}{\t{}} \\ \subtype{\t3}{\t{}}
        \\\\
        \p{}={\combeff{\p1}{\p2}{\p3}}
        }
        {\ghastyeff{\cond{\e1}{\e2}{\e3}}{\t{}}{\p{}}}
\]
\end{center}

\caption{Primary Typing Rules}
\label{f:type}
\end{figure*}

\begin{figure}[t!]
\begin{center}
\begin{tabular}{ll}

\combeff{\p{}'}{\p{}}{\p{}} = \p{} 
 & $\dt{\addone} = \proctype{\num}{\num}{\noeffect}$ \\
$\combeff{\t{\x{}}}{\tt}{\s{\x{}}} = (\usym\ \t{}\ \s{})_{\x{}}$
 & $\dt{\nott} = \proctype{\top}{\bool}{\noeffect}$
\\
\combeff{\tt}{\p1}{\p2} = \p1 
 & $\dt{\procp} = \proctype{\top}{\bool}{\proctype{\scheme|bot|}{\top}{\noeffect}} $\\
\combeff{\ff}{\p1}{\p2} = \p2 
 & \dt{\numberp} = \proctype{\top}{\bool}{\num} \\
\combeff{\p{}}{\tt}{\ff} = \p{} 
 & \dt{\boolp} = \proctype{\top}{\bool}{\bool}\\
\combeff{\p1}{\p2}{\p3} = \noeffect & \\
\end{tabular}
\end{center}
\caption{Auxiliary Operations}
\label{f:aux}
\end{figure}

Figure \ref{f:syntax} specifies the syntax of \lts~ programs. An expression
is either a value, a variable, an application, or a conditional.  The set of
values consists of abstractions, numbers, booleans, and constants.  Binding
occurrences of variables are explicitly annotated with types.  Types are
either $\top$, function types, base types, or unions of some finite
collection of types. We refer to the decorations on function types as {\em
\lateff{}s} and explain them, along with {\em \lexeff{}s}, below in
conjunction with the typing rules.  For brevity, we abbreviate
$(\usym\, \tt\, \ff)$ as \bool and $(\usym)$ as $\bot$.

\begin{figure}[t!]
\begin{center}
\begin{tabular}{ll}
 \G{} + \t{\x{}}
= \G{}[\hastype{\x{}}{\restrict{\G{}(x)}{\t{}}}]
\\
 \G{} + \x{} = \G{}[\hastype{\x{}}{\remove{\G{}(x)}{\ff}}]
\\
\G{} + \noeffect = \G{}
\\
\G{} - \t{\x{}} = \G{}[\hastype{\x{}}{\remove{\G{}(x)}{\t{}}}]
\\
\G{} - \x{} = \G{}[\hastype{\x{}}{\ff}]
\\
 \G{} - \noeffect = \G{}
\end{tabular}
\begin{tabular}{ll}
 \restrict{\s{}}{\t{}} = \s{} \mbox{\ when $\subtype{\s{}}{\t{}}$}
\\ \restrict{\s{}}{(\usym\ \t{} \ldots)} =
(\usym\ \restrict{\s{}}{\t{}} \ldots)
\\ \restrict{\s{}}{\t{}} = \t{} \mbox{\ otherwise}
\\ \remove{\s{}}{\t{}} = $\bot$ \mbox{\ when $\subtype{\s{}}{\t{}}$}
\\ \remove{\s{}}{(\usym\ \t{} \ldots)} =
(\usym\ \remove{\s{}}{\t{}} \ldots)
\\ \remove{\s{}}{\t{}} = \s{} \mbox{\ otherwise}
\end{tabular}
\end{center}
\caption{Environment Operations}
\label{f:envop}
\end{figure}


The operational semantics is standard: see figure~\ref{f:op}.
Following Scheme and Lisp tradition, any non-\ff value is treated
as \tt.

\subsection{Preliminaries}

The key feature of \lts\ is its support for assigning distinct types
to distinct occurrences of a variable based on control flow
criteria. For example, to type the expression
\begin{schemedisplay}
(lambda (x : (U Number Boolean)) 
  (if (number? x) (= x 1) (not x)))
\end{schemedisplay}
the type system must use \scheme{Number} for \scheme|x| in the \emph{then} branch of the
conditional and \scheme{Boolean} in the \emph{else} branch. If it can
distinguish these occurrences and project out the proper component of the
declared type \scheme{(U Number Boolean)},  the computed type of the
function is 
$$\proctype{(\usym\ \num\ \bool)}{\bool}{} . $$

The type system for \lts~shows how to distinguish these occurrences;
 its presentation consists
of two parts.  The first are those rules that the programmer must know
and that are used in the implementation of Typed Scheme.  The second
set of rules are needed only to establish type soundness; 
 these rules are unnecessary outside of the proof of the
main theorem.

{\bf Visible Predicates} 
Judgments of \lts\ involve both types and visible predicates (see the
production for $\p{}$ in figure \ref{f:syntax}). The former are
standard. The latter are used to
accumulate information about expressions that affect the flow of
control and thus demand a split for different branches of a
conditional. Of course, a syntactic match would help little, because
programmers of scripts tend to write their own predicates and compose
logical expressions with combinators. Also, programmer-defined
datatypes extend the set of predicates.

{\bf Latent Predicates}
In order to accommodate programmer-defined functions that are used as
predicates, the type system of \lts\ uses {\lateff}{s}\ (see \ph{} in
figure \ref{f:syntax}) to annotate function types. Syntactically
speaking, a \lateff\ is a single type $\phi$ atop the arrow-type
constructor that identifies the function as a predicate for
$\phi$. This \lateff-annotation allows a uniform treatment of built-in
 and user-defined predicates. For example,
$$ \numberp :  {\proctype {\top} {\bool} {\num}} $$
says that \numberp\ is a discriminator for numbers.
  An eta-expansion preserves this property: 
 $$ (\lambda\ (x \mathbf{:} \top)\ (\numberp\ x)) :  {\proctype {\top} {\bool} {\num}} . $$

Thus far, \emph{higher-order} {\lateff}s are useful in just one case: \scheme{procedure?}. For
uniformity, the syntax accommodates the general case. We intend to study an
integration of \lateff{}s with higher-order
contracts~\cite{ff:ho-contracts} and expect to find additional uses.

The \lts\ calculus also accommodates logical combinations of predicates. Thus, if
 a program contains a test expression such as:
\begin{schemedisplay}
(if (number? x) #t (boolean? x))
\end{schemedisplay}
then Typed Scheme computes the appropriate \lexeff\ for this union, which is $(\usym\, \num\, \bool)_{\x{}}$.  This
 information is propagated so that a programmer-defined function receives a
 corresponding \lateff. That is, the \scheme|bool-or-number| function:
\begin{schemedisplay}
(lambda (x : Any) (if (number? x) #t (boolean? x)))
\end{schemedisplay}
 acts like a predicate of type \scheme|(Any -U> Boolean)| and is used to split types
 in different branches of a conditional.

\subsection{Typing Rules}
Equipped with types and predicates, we turn to the
typing rules. They derive judgements of the form
$${\ghastyeff{\e{}}{\t{}}{\p{}}} . $$ It states that in type
environment $\Gamma$, expression \e{} has type \t{} and \lexeff\ \p{}.
The latter is used to change the type environment in conjunction
with \scheme{if} expressions.\footnote{Other control flow constructs in
Scheme are almost always macros that expand into \scheme|if|, and that
the typechecker can properly check.}  The
type system proper comprises the ten rules in figure \ref{f:type}.

The rule {\sc T-If} is the key part of the system, and shows
how \emph{visible} predicates are treated.  To
accommodate Scheme style, we allow expressions with {\em any\/}
type as tests. Most importantly, though, the rule uses the \lexeff\ of
the test to modify the type environment for the verification of the
types in the two conditional branches.  When a variable is used as the
test, we know that it cannot be \ff in the {\it then} branch, and
must be in the {\it else} branch.

While many of the type-checking rules appear familiar,
the presence of \lexeff\ distinguishes them from ordinary rules:
\begin{itemize}
\item {\sc T-Var} assigns a variable its type from the type environment and
names the variable itself as the \lexeff.

\item Boolean constants have Boolean type and a \lexeff\ that depends on
their truth value. Since numbers and primitive functions are always treated as true values,
they have visible predicate \tt. 

\item
When we abstract over a predicate, the abstraction should reflect the
test being performed.  This is accomplished with the {\sc T-AbsPred}
rule, which gives an abstraction a latent predicate if the body of the
abstraction has a visible predicate referring to the
abstracted variable, as in the \scheme|bool-or-number| example.

Otherwise, abstractions have their usual type; the visible predicate of their body is ignored.  The
  visible predicate of an abstraction is \tt, since abstractions are
  treated that way by \scheme|if|.  

\item Checking plain applications proceeds as normal. The antecedents include \lateff s and
\lexeff s but those are ignored in the consequent. 

\item The {\sc T-AppPred} rule shows how the type system exploits latent predicates.  The
application of a function with \lateff\ to a variable turns the latent
predicate into a visible predicate on the variable (\s{\x{}}). The proper
interpretation of this \lexeff\ is that the application produces \tt if
and only if \x{} has a value of type \s{}.
\end{itemize}

\begin{figure}[t!]
\[
\inferrule[S-Refl]{}{\subtype{\t{}}{\t{}}}
\qquad
\inferrule[S-Fun]{\subtype{\s1}{\t1} \\ \subtype{\t2}{\s2} \\\\ \mbox{$\ph{} = \ph{}'$\ or\ $\ph{}' = \noeffect$ }}
        {\subtype{\proctype{\t1}{\t2}{\ph{}}}{\proctype{\s1}{\s2}{\ph{}'}}}
\]

\[
\inferrule[S-UnionSuper]{\subtype{\t{}}{\s{i}} \\ 1 \leq i \leq n}
        {\subtype{\t{}}{(\usym\ \s1 \cdots \s{n})}}
\qquad
\inferrule[S-UnionSub]{\subtype{\t{i}}{\s{}} \mbox{\ for all $1 \leq i \leq n$}}
        {\subtype{(\usym\ \t1 \cdots \t{n})}{\s{}}}
\]
\caption{Subtyping Relation}
\label{f:sub}
\end{figure}

\begin{figure}[htbp]
\[
\inferrule[T-AppPredTrue]{\ghastyeff{\e1}{\t{}'}{\p{}} \\ \ghastyeff{\e2}{\t{}}{\p{}'}
  \\\\ \subtype{\t{}}{\t0} \\ \subtype{\t{}}{\s{}}
\\
{\subtype{\t{}'}{\proctype{\t0}{\t1}{\s{}}}}}
        {\ghastyeff{\comb{\e1}{\e2}}{\t1}{\tt}}
\qquad\qquad
\inferrule[T-AppPredFalse]{\ghastyeff{\e1}{\t{}'}{\p{}} \\ \ghastyeff{\vv{}}{\t{}}{\p{}'}
  \\\\ \subtype{\t{}}{\t0} \\ \notsubtype{\t{}}{\s{}} \\ \vv{} \ \mathrm{closed}
\\\\
{\subtype{\t{}'}{\proctype{\t0}{\t1}{\s{}}}}}
        {\ghastyeff{\comb{\e1}{\vv{}}}{\t1}{\ff}}
\]

\[
\inferrule[T-IfTrue]
        {\ghastyeff{\e1}{\t1}{\tt}
        \\
        \hastyeff{\G{}}{\e2}{\t2}{\p2}
        \\\\
        \subtype{\t2}{\t{}}
        }
        {\ghasty{\cond{\e1}{\e2}{\e3}}{\t{}}}
\qquad\qquad
\inferrule[T-IfFalse]
        {\ghastyeff{\e1}{\t1}{\ff}
        \\
        \hastyeff{\G{}}{\e3}{\t3}{\p3}
        \\\\
        \subtype{\t3}{\t{}}
        }
        {\ghasty{\cond{\e1}{\e2}{\e3}}{\t{}}}
\]

\[
\inferrule[SE-Refl]{}{\subeff{\p{}}{\p{}}}
\qquad\qquad
\inferrule[SE-None]{}{\subeff{\p{}}{\noeffect}}
\]

\[
\inferrule[SE-True]{\p{} \neq \ff}{\subeff{\tt}{\p{}}}
\qquad\qquad
\inferrule[SE-False]{\p{} \neq \tt}{\subeff{\ff}{\p{}}}
\]
\caption{Auxiliary Typing Rules}
\label{f:type-aux}
\end{figure}

\begin{figure}[htbp]
\[
\inferrule*[lab={E-Delta}]
        {\del{\c{}}{\vv{}} = \vv{}'}
        {\step{\comb{\c{}}{\vv{}}}{\vv{}'}}
\qquad\qquad
\inferrule[E-Beta]{}
      {\step{\comb{\abs{\x{}}{\t{}}{\e{}}}{\vv{}}}{\subs{\e{}}{\x{}}{\vv{}}}}
\]

\[
\inferrule[E-IfFalse]{\vv{} = \ff}
      {\step{\cond{\vv{}}{\e2}{\e3}}{\e3}}
\qquad\qquad
\inferrule[E-IfTrue]{\vv{} \neq \ff}
      {\step{\cond{\vv{}}{\e2}{\e3}}{\e2}}
\]

\[
\inferrule*
        {\step{L}{R}}
        {\reduce{\E{}[L]}{\E{}[R]}}
\]
$$
\del{\addone}{\n{}} = \n{} + 1
$$
$$
\del{\nott}{\ff} = \tt
\qquad
\del{\nott}{\vv{}} = \ff \mbox{\ $\vv{} \neq \ff$}
$$
$$
\del{\numberp}{\n{}} = \tt
\qquad
\del{\numberp}{\vv{}} = \ff
$$
$$
\del{\boolp}{\b{}} = \tt
\qquad
\del{\boolp}{\vv{}} = \ff
$$
$$
\del{\procp}{\abs{\x{}}{\t{}}{\e{}}} = \tt
\qquad
\del{\procp}{\c{}} = \tt
$$
$$
\del{\procp}{\vv{}} = \ff \mbox{\ otherwise}
$$
\caption{Operational Semantics}
\label{f:op}
\end{figure}

Figure \ref{f:aux} defines a number of auxiliary typing operations.
The mapping from constants to types is standard.  The ternary \combeff{-}{-}{-}
metafunction combines the effects of the test, then and else branches
of an \scheme|if| expression.  The most interesting case is the second
one, which
handles expressions such as this:
\begin{schemedisplay}
(if (number? x) #t (boolean? x))
\end{schemedisplay}
\noindent the equivalent of an \scheme|or| expression.  The combined effect
is $(\usym \num \bool)_{\x{}}$, as expected.

The environment operations, specified in figure \ref{f:envop}, combine
a \lexeff\ with a type environment, updating the type of the
appropriate variable.  Thus, \restrict{\s{}}{\t{}}
 is \s{} restricted to be a subtype of \t{}, and \remove{\s{}}{\t{}}
is \s{} without the portions that are subtypes of \t{}.  The only
non-trivial cases are for union types.

For the motivating example from the beginning of this section,

\begin{schemedisplay}
(lambda (x : (U Number Boolean)) (if (number? x) (= x 1) (not x)))
\end{schemedisplay}
\noindent
we can
now see that the test of the \scheme{if} expression has type \bool\
and \lexeff\ $\num_{\x{}}$.  As a consequence, the
\emph{then} branch is type-checked in an environment where \x{} has type
\num; in the \emph{else} branch, \x{} is assigned \bool. 

\paragraph{Subtyping}
The definition of subtyping is given in figure \ref{f:sub}.  The rules
are for the most part standard, with the rules for union types adapted
from  Pierce's~\cite{pierce-union}.  
One important consequence of these rules is that \scheme|bot| is below
all other types.  This type is useful for typing functions that do
not return, as well as for
defining a supertype of all function types.

We do not include a transitivity rule for the subtyping relation, but
instead prove that the subtyping relation as given is transitive.
This choice simplifies the proof in a few key places.

The rules for subtyping allow function types with {\lateff}s to be
used in a context that expects a function that is not a predicate.  This
is especially important for \scheme|procedure?|, which handles
functions regardless of \lateff.

\subsection{Proof-Technical Typing Rules}

The typing rules in figure \ref{f:type} do not suffice for the 
soundness proof.  
To see why, consider the function from above, applied to the
argument \scheme|#f|.  By the {\sc E-Beta} rule, this reduces to

\begin{schemedisplay}
(if (number? #f) (= #f 1) (not #f))
\end{schemedisplay}
Unfortunately, this program is not well-typed according the primary
typing rules, since \scheme|=| requires numeric arguments.  Of course,
this program reduces in just a few steps to \scheme|#t|, which is an
appropriate value for the original type.  To prove type soundness
in the style of Wright and Felleisen~\cite{wf:type-soundness}, however, every intermediate term must be
typeable.  So our types system must know to ignore the \emph{then} branch of our reduced term.  

To this end, we extend the type system with the rules in figure~\ref{f:type-aux}.
This extension assigns the desired type to our reduced expression, because
\scheme|(number? #f)| has \lexeff\ \ff.  Put differently, we can disregard the
\emph{then} branch, using rule {\sc T-IfFalse}.\footnote{The rules in
figure 6 are similar to rules used for the same purpose in systems
with a \texttt{typecase} construct, such as \citet{weirich98}.}

In order to properly state the subject reduction lemma, we need to
relate the \lexeff s of terms in a reduction sequence.  To this end, we
define a \subpred\ relation, written $\subeff{\p{}}{\p{}'}$.
  The relation is defined in figure \ref{f:type-aux}; it
 is not used in the subtyping or typing rules,
being  needed only for the soundness proof.

We can now prove the traditional lemmas.  We work only with closed terms, since it simplifies the possible predicates of the expression.

\begin{lemma}[Preservation]
  If \mthastyeff{\e{}}{\t{}}{\p{}} (with \e{} closed) and \reduce{\e{}}{\e{}'}, then
  \mthastyeff{\e{}'}{\t{}'}{\p{}'} where \subtype{\t{}'}{\t{}} and \subeff{\p{}'}{\p{}}.
\end{lemma}

{\it  Proof Sketch} This is a corollary of two other lemmas: that
plugging a well typed term into the hole of an evaluation preserves
the type of the resulting term, and that the $\step{\e1}{\e2}$
preserves type when \e1 is closed.  These two lemmas are both proved
by induction on the relevant typing derivations.  \qed

\begin{lemma}[Progress]
  If \mthastyeff{\e{}}{\t{}}{\p{}} (with \e{} closed) then either \e{}
  is a value or \reduce{\e{}}{\e{}'} for some $\e{}'$.
\end{lemma}

{\it  Proof Sketch}
By induction on the derivation of \ghastyeff{\e{}}{\t{}}{\p{}}.  
\qed

From these, soundness for the extended type system follows.
Programs with untypable subexpressions, however, are not useful in real
programs. We only needed to consider them, as well as our
additional rules, for our proof of soundness. Fortunately, we can also
show that the additional, proof-theoretic, rules are needed only for
the type soundness proof, not the result. Therefore, we obtain
the desired type soundness result.

\begin{theorem}[Soundness]
  If \ghastyeff{\e{}}{\t{}}{\p{}}, with \e{} closed, using only the rules in figure~\ref{f:type},
  and \t{} is a base type, one of the following holds 
  \begin{enumerate}
    \item \e{} reduces forever, or
    \item \reduces{\e{}}{\vv{}} where \hastyeff{\
}{\vv{}}{\s{}}{\p{}'} and  \subtype{\s{}}{\t{}} and \subeff{\p{}'}{\p{}}.
  \end{enumerate}
\end{theorem}

{\it  Proof Sketch}
First, this is a  corollary of soundness if the
requirement is only that \vv{} typechecks in the extended system, since
it types strictly more terms. Second, the extended system
agrees with the non-extended system on all values of ground type 
(numbers and booleans).  Thus, \vv{} has the appropriate type even in
the original system.
\qed

\label{sec:soundness}




\subsection{Mechanized Support}

We employed two mechanical systems for the exploration of the model and
 the proof of the soundness theorem: Isabelle/HOL~\cite{isabelle}
 and PLT Redex~\cite{jfff:redex}. Indeed, we freely moved back and forth
 between the two, and without doing so, we would not have been able to
 formalize the type system and verify its soundness in an adequate and
 timely manner.

For the proof of type soundness, we used Isabelle/HOL together with
the nominal-isabelle package~\cite{nominal-isabelle}.  Expressing a
type system in Isabelle/HOL is almost as easy as writing down the
typing rules of figures~\ref{f:type} and~\ref{f:type-aux} (our
formalization runs to ~5000 lines). To represent the reduction
semantics (from figure~\ref{f:op}) we turn evaluation contexts into
functions from expressions to expressions, which makes it relatively
straightforward to state and prove lemmas about the connection between
the type system and the semantics. Unfortunately, this design choice
prevents us from evaluating sample programs in Isabelle/HOL, which is
especially important when a proof attempt fails.

Since we experienced such failures, we also used the PLT Redex
system~\cite{jfff:redex} to explore the semantics and the type system
of Typed Scheme. PLT Redex programmers can write down a reduction
semantics as easily as Isabelle/HOL programmers can write down typing
rules. That is, each line in figures~\ref{f:syntax} and~\ref{f:op}
corresponds to one line in a Redex model.  Our entire Redex model,
with examples, is less than 500 lines.  Redex comes with visualization
tools for exploring the reduction of individual programs in the object
language. In support of subject reduction proofs, language designers
can request the execution of a predicate for each ``node'' in the
reduction sequences (or graphs). Nodes and transitions that violate a
subject reduction property are painted in distinct colors,
facilitating example-based exploration of type soundness proofs.

Every time we were stuck in our Isabelle/HOL proof, we would turn to Redex
 to develop more intuition about the type system and semantics. We would
 then change the type system of the Redex model until the violations of
 subject reduction disappeared. At that point, we would translate the
 changes in the Redex model into changes in our Isabelle/HOL model and
 restart our proof attempt. Switching back and forth in this manner helped
 us improve the primary typing rules and determine the shape of the
 auxiliary typing rules in figure~\ref{f:type-aux}. Once we had those,
 pushing the proof through Isabelle/HOL was a labor-intensive mechanization
 of the standard proof technique for type soundness.







\section{Formalizing Refinements}
\label{sec:refinement}
\newcommand{\rc}[1]{\ma{\R{\c{}}{#1}}}
\newcommand{\rct}{\rc{\t{}}}

It is straightforward to add refinement types to the \lts\ calculus.
We extend the grammar with the new type constructor \rct, which
is the refinement defined by the built-in function \c{}, which has
argument type \t{}.\footnote{\t{} is inferred from the type of \c{} in
the implemented system, as demonstrated in section~\ref{sec:example}.}
  We restrict refinements to built-in functions
so that refinement types can be given to closed expressions and values
such as \scheme|0|.  We then add two new constants, \scheme|even?|, with type
$$\proctype{\num}{\bool}{\R{\evenp}{\num}}$$
\noindent
 and
\scheme|odd?|, with type
$$\proctype{\num}{\bool}{\R{\oddp}{\num}}$$
\noindent
and the obvious semantics.

The subtyping rules for refinements require an additional environment $\Delta$,
which specifies which built-ins may be used as refinements.   Extending the
existing subtyping rules with this environment is straightforward,
giving a new judgement of the form \dsubtype{\t1}{\t2}, with the
subscript $r$ distinguishing this judgement from the earlier
subtyping judgement.  As an example, the extended version of the {\sc S-Refl} rule is
$$
\inferrule{}{\dsubtype{\t{}}{\t{}}}
$$
\noindent
The new rule for refinement types is

\[
\inferrule[]{\c{} \in \Delta
\\
\dt{\c{}} = \proctype{\t1}{\t2}{\ph{}} \\
\dsubtype{\t1}{\t{}}
}{
\dsubtype{\R{\c{}}{\t1}}{\t{}}}
\]
\noindent
This rule states that a refinement of type \t1 is a subtype of any
type of which \t1 is a subtype.  As expected, this means that
\dsubtype{\rct}{\t{}}. 

The addition of this environment to the subtyping judgement requires a
similar addition to the typing judgement, which now has the form
\dghastyeff{\e{}}{\t{}}{\p{}}.  

This subtyping rule, along with the constants \scheme|even?| and
\scheme|odd?|, are sufficient to write useful examples.  For
example, this function consumes an even-consuming function and a
number, and uses the function if and only if the number is even.

\begin{schemedisplay}
(lambda ([f : ((Refinement even? Number) -> Number)] [n : Number])
  (if (even? n) (f n) n))
\end{schemedisplay}

No additional type rules are necessary for this extension.
Additionally, any expression of type \rct can be used as if it has
type \t{}, meaning that standard arithmetic operations still work on
even and odd numbers.  

\subsection{Soundness}

Proving soundness for the extended system with refinements raises the
interesting question of what additional errors are prevented by the
refinement type extension.  The answer is none; no additional
behavior is ruled out.  This is unsurprising, of course, since the
soundness theorem from section~\ref{sec:soundness} does not allow the
possibility of any errors.  But even if errors were added to the
operational semantics, such as division by zero, none of these errors
would be prevented by the refinement type system.  Instead, refinement
types allow the specification and enforcement of types that
 do not have any necessary correspondence to the operational
semantics of the language.  

We therefore adopt a different proof strategy.  Specifically, we erase
the refinement types 
and are left with a typeable term, which reduces appropriately.
  Given a type in the extended
language, we can compute a type without refinement types, simply by
erasing all occurrences of \R{\c{}}{\t{}} to \t{}.  The definition
of this function, $\mathsf{erase}_{\t{}}$ is given in
figure~\ref{fig:erase}, along with its extension to terms
($\mathsf{erase}_{\lambda}$), predicates ($\mathsf{erase}_{\p{}}$),
 environments ($\mathsf{erase}_{\Gamma}$) and judgments
 ($\mathsf{erase}_{\,\vdash}$).
  We also assume the obvious modifications to
$\delta_{\t{}}$.

\begin{figure}
\[
\begin{tabular}{c}
$
\begin{array}{l@{\quad =\quad}l}
\er{\rct} & \er{\t{}} \\
\er{\proctype{\t{}}{\s{}}{\t{}'}} &
\proctype{\er{\t{}}}{\er{\s{}}}{\er{\t{}'}} \\
\er{\proctype{\t{}}{\s{}}{\bullet}} &
\proctype{\er{\t{}}}{\er{\s{}}}{\bullet} \\
\er{\num} & \num \\
\er{\tt} & \tt \\
\er{\ff} & \ff \\
\er{\top} & \top \\
\er{(\usym\ \t{} \dots)} & (\usym\ \er{\t{}} \dots) \vspace{2mm}\\

\erl{\abs{\x{}}{\t{}}{\e{}}} & \abs{\x{}}{\er{\t{}}}{\erl{\e{}}} \\
\erl{\comb{\e1}{\e2}} & \comb{\erl{\e1}}{\erl{\e2}}\\
\erl{\cond{\e1}{\e2}{\e3}} & \cond{\erl{\e1}}{\erl{\e2}}{\erl{\e3}}\\
\erl{\n{}} & \n{} \\
\erl{\c{}} & \c{} \\
\erl{\b{}} & \b{} \\
\erl{\x{}} & \x{} \vspace{2mm}\\

\erp{\t{\x{}}} & \er{\t{}}_{\x{}} \\
\erp{\x{}} & \x{} \\
\erp{\bullet{}} & \bullet{} \\
\erp{\ttt{}} & \ttt{} \\
\erp{\fft{}} & \fft{} \vspace{2mm}\\

\erg{{\x{} : \t{}}, \ldots} & {\x{} : \er{\t{}}},
\ldots \vspace{2mm}\\

\erj{\ghastyeff{\e{}}{\t{}}{\p{}}} & \hastyeff{\erg{\Gamma}}{\erl{\e{}}}{\er{\t{}}}{\erp{\p{}}}
\end{array}
$
\end{tabular}
\]
\caption{Erasure Metafunctions}
\label{fig:erase}
\end{figure}

With these definitions in hand, we can conclude the necessary lemmas
for proving soundness.  

\begin{lemma}[Typing Erased Terms]
If \dghastyeff{\e{}}{\t{}}{\p{}}, then
\erj{\ghastyeff{{\e{}}}{{\t{}}}{{\p{}}}}. 
\end{lemma}

\begin{proof}
By induction on the derivation of \dghastyeff{\e{}}{\t{}}{\p{}}. \qed
\end{proof}

\begin{lemma}[Reducing Erased Terms]
If \reduce{\e1}{\e2}, then \reduce{\erl{\e1}}{\erl{\e2}}.
\end{lemma}

\begin{proof}
By induction on the derivation of \reduce{\e1}{\e2}. \qed
\end{proof}

We can combine these lemmas with our earlier preservation and progress
lemmas to conclude soundness.  

\begin{theorem}[Soundness with Refinement Types]
  If \dghastyeff{\e{}}{\t{}}{\p{}}, with \e{} closed, using only the rules in figure~\ref{f:type},
  and \t{} is a base type or a refinement of a base type, one of the following holds 
  \begin{enumerate}
    \item \e{} reduces forever, or
    \item \reduces{\e{}}{\vv{}} where \erj{\hastyeff{\
}{{\vv{}}}{{\s{}}}{{\p{}'}}} and  \subtype{\er{\s{}}}{\er{\t{}}} and \subeff{\erp{\p{}'}}{\erp{\p{}}}.
  \end{enumerate}
\end{theorem}


\section{From \lts~To Typed Scheme}

It is easy to design a type system, and it is
reasonably straightforward to validate some theoretical
property. However, the
true proof of a type system is a pragmatic evaluation. To this end, it
is imperative to integrate the novel ideas with an existing
programming language. Otherwise it is difficult to demonstrate that
the type system accommodates the kind of programming style that people
find natural and that it serves its intended purpose.

To evaluate occurrence typing rigorously, we have implemented Typed
 Scheme. Naturally, occurrence typing with refinements, in the spirit of 
 \lts\ makes up only the core of this language;
  we have supplemented it with a number of important ingredients,
 both at the level of types and at the level of large-scale
 programming. 

\subsection{Type System Extensions}

As argued in the introduction, Scheme programmers borrow a number of ideas
 from type systems to reason about their programs. Chief among them is
 parametric polymorphism.  Typed Scheme therefore allows programmers to
 define and use polymorphic functions.  For example, the
 \scheme|map| function is defined as follows:

\begin{schemedisplay}
(define: (a b) (map [f : (a -> b)] [l : (Listof a)]) : (Listof b)
  (if (null? l) l
      (cons (f (car l)) (map f (cdr l)))))
\end{schemedisplay}
\noindent
The definition explicitly quantifies over type variables \scheme|a|
and \scheme|b| and then uses these variables in the type signature.
The body of the definition, however, is identical to the one for
untyped \scheme|map|; in particular, no type application is required
for the recursive call to \scheme|map|.  Instead, the type system
infers appropriate instantiations for \scheme|a| and \scheme|b| for
the recursive call.

In addition to parametric polymorphism, Scheme programmers also exploit
recursive subtypes of S-expressions to encode a wide range of information
as data. To support arbitrary regular types over S-expressions as well as
 conventional structures, Typed Scheme provides explicit recursive
types, though the programmer need not manually fold and unfold instances of
these types. 

Consider the type of binary trees over \scheme|cons| cells:
\begin{schemedisplay}
(define-type-alias STree (mu t (U Number (cons t t))))
\end{schemedisplay}
A function for summing the leaves of such a tree is straightforward: 
\begin{schemedisplay}
(define: (sum-tree [s : STree]) : Number
  (cond [(number? s) s]
	[else (+ (sum-tree (car s)) (sum-tree (cdr s)))]))
\end{schemedisplay}
In this function, occurrence typing allows us to discriminate
between the different branches of the union; the (un)folding of the
recursive (tree) type happens automatically.

Finally, Typed Scheme supports a rich set of base types, including vectors,
boxes, parameters, ports, and many others. It also provides type aliasing,  which
 greatly facilitates type readability.  


\subsection{Local Type Inference}

In order to further relieve the annotation burden on programmers,
Typed Scheme provides two simple instances of what has been called ``local''
type inference \cite{pierce:lti}.\footnote{This modicum of inference
  is similar to that in recent releases of Java~\cite{jls3}.}  First,
local non-recursive bindings do not require type annotations.  For example,
the following fragment typechecks without annotations on the local bindings:

\begin{schemedisplay}
(define: (m [z : Number]) : Number
  (let* ([x z]
         [y (* x x)])
    (- y 1)))
\end{schemedisplay}

\noindent
By examining the right-hand sides of the \scheme|let*|, the
typechecker can determine that both \scheme|x| and \scheme|y| should
have type \scheme|Number|.  

The use of internal definitions can complicate this inference process.
For example, the above code could be written as follows:

\begin{schemedisplay}
(define: (m [z : Number]) : Number
  (define x z)
  (define y (* x x))
  (- y 1))
\end{schemedisplay}
  
This fragment is macro-expanded into a \scheme|letrec|; however, recursive
binding is not required for typechecking this code.  Therefore, the
typechecker analyzes the \scheme|letrec| expression and determines if
all of the bindings can be treated non-recursively.  If so, the above
inference method is applied.

Second, local inference also allows the type arguments to polymorphic
functions to be omitted.  For example, the following use of
\scheme|map| does not require explicit type instantiation:  

\begin{schemedisplay}
(map (lambda: ([x : Number]) (+ x 1)) '(1 2 3))
\end{schemedisplay}
\noindent
To accommodate this form of inference, the typechecker first determines the type of the
argument expressions, in this case \scheme|(Number -> Number)| and
\scheme|(Listof Number)|, as well as the operator, here
\scheme|(All (a b) ((a -> b) (Listof a) -> (Listof b)))|.  Then it
matches the argument types against the body of the operator type,
generating a substitution.  Finally, the substitution is applied to
the function result type to determine the type of the entire
expression.  

For cases such as the above, this process is quite straightforward.  
When subtyping is involved, however, the process is complex.  Consider
this, seemingly similar, example:

\begin{schemedisplay}
(map (lambda: ([x : Any]) x) '(1 2 3))
\end{schemedisplay}
\noindent
Again, the second operand has type \scheme|(Listof Number)|,
suggesting that \scheme|map|'s type variable \scheme|b| should substituted with
\scheme|Number|, the first operand has type  \scheme|(Any ->
Any)|, suggesting that both \scheme|a| and \scheme|b| should be
\scheme|Any|.  The solution is to find a common supertype of
\scheme|Number| and \scheme|Any|, and use that to substitute for
\scheme|a|.  

Unfortunately, this process does not always succeed.  Therefore, the
programmer must sometimes annotate the arguments or the function to
enable the typechecker to find the correct substitution.  For example,
this annotation instantiates \scheme|foldl| at \scheme|Number| and
\scheme|Any|:

\vspace{2mm}
\noindent
\#\{{\it foldl} {\bf @ Number Any}\}
\vspace{2mm}

\noindent
In practice, we have rarely needed these annotations; local inference almost always succeeds. 

\subsection{Adapting Scheme Features}

PLT Scheme comes with numerous
constructs that need explicit support from the type system.  We
describe several of the more important ones here.

\begin{itemize}

\item The most important one is the \scheme{structure} system.  A
\scheme{define-struct} definition is {\em the\/} fundamental method for
constructing new varieties of data in PLT Scheme.  This form of definition introduces
constructors, predicates, field selectors, and field mutators.  Typed
Scheme includes a matching \scheme|define-struct:| form.  Thus the
untyped definition 

\leftcodeskip\parindent
\rightcodeskip 1pt
\begin{schemedisplay}
(define-struct A (x y))
\end{schemedisplay}

which defines a structure \scheme|A|, with fields \scheme|x| and
\scheme|y|, becomes the following in Typed Scheme: 

\begin{schemedisplay}
(define-struct: A ([x : Number] [y : String]))
\end{schemedisplay}

Unsurprisingly, all fields have type annotations.

The \scheme|define-struct:| form, like \scheme|define-struct|,
introduces the predicate \scheme|A?|. Scheme programmers use this
predicate to discriminate instances of \scheme{A} from other values,
and the occurrence typing system must therefore be aware of it. The
\scheme{define-struct:} definition facility can also
automatically introduce recursive types, similar to those introduced
via ML's datatype construct.

Programmers may define structures as extensions of an existing
structure, similar to extensions of classes in object-oriented
languages. An extended structure inherits all the fields of its parent
structure. Furthermore, its parent predicate cannot discriminate instances
of the parent structure from instances of the child structure.  Hence, it
is imperative to integrate structures with the type system at a fundamental level.

\item PLT Scheme encourages placing all code in modules, but the top
  level still provides valuable interactivity.  Typed Scheme supports
  both definitions and expression at the top-level, but support is
  necessarily limited by the restrictions of typechecking a form at a
  time.  For example, mutually recursive top-level functions 
  cannot be defined, since type checking of the first happens before
  the second is entered.  

\item Variable-arity functions also demand special attention from the type
perspective. PLT Scheme supports two forms of variable-arity functions: 
rest parameters, which bundle up extra arguments into a list; and
\scheme|case-lambda|~\cite{dybvig-case-lambda}, which, roughly speaking,
introduces dynamic overloading by arity.  A careful adaptation of
the solutions employed for mainstream languages such as Java and C\#
suffices for some of these features; for others, we have developed
additional type 
system extensions to handle the unique features of PLT
Scheme~\cite{sthf:variable-arity}. 

\item Dually, Scheme supports multiple-value returns, meaning a procedure
may return multiple values simultaneously without first bundling them up in
a tuple (or other compound values).  Multiple values are given special
treatment in the type checker because the construct for returning multiple
values is a primitive function (\scheme|values|), which can be used in
higher-order contexts.  Such higher-order uses of \scheme|values|
benefit from extensions to handle variable-arity polymorphism, as
described above~\cite{sthf:variable-arity}.

\item Finally, Scheme programmers use the \scheme|apply|
function, especially in conjunction with variable-arity functions. The
\scheme{apply} function consumes a function, a number of values, plus a
list of additional values; it then applies the function to all these
values. 

Because of its use in conjunction with variable-arity functions, we
type-check the application of \scheme|apply| specially and allow its
 use with variable-arity functions of the appropriate type.


For example, the common Scheme idiom of \scheme|apply|ing the
function \scheme|+| to a list of numbers to sum them works in Typed
Scheme: \scheme{(apply + (list 1 2 3 4))}.
\end{itemize}


\subsection{Special Scheme Functions}

A number of Scheme functions, either because of their special semantics or
their particular roles in the reasoning process of Scheme programmers, are
assigned types that demand some explanation. Here we cover just two interesting examples:
\scheme{filter} and \scheme{call/cc}. 

An important Scheme function, as we saw in section
\ref{s:overview}, is \scheme|filter|.  

When \scheme|filter| is used with predicate \scheme|p?|, the programmer
knows that every element of the resulting list satisfies \scheme|p?|.
The type system should have this
knowledge as well, and in Typed Scheme it does:

\begin{schemedisplay}
filter : (All (a b) ((a bover Boolean) (Listof a) -> (Listof b))
\end{schemedisplay}

\noindent
Here we write \scheme|(a bover Boolean)| for the type of functions from
\scheme|a| to \scheme|Boolean| that are predicates for type \scheme|b|.
Note how the \lateff\ of filter becomes the type of the resulting
elements. In a setting without occurrence typing, this effect has only
been achieved with dependent types or with explicit casting operations. 

For an example, consider the following definition:
\begin{schemedisplay}
(define: the-numbers (Listof Number)
  (let ([lst (list 'a 1 'b 2 'c 3)])
    (map add1 (filter number? lst))))
\end{schemedisplay}
Here \scheme{the-numbers} has type \scheme{(Listof Number)}
even though it is the result of filtering numbers from a list that contains
both symbols and numbers. Using Typed Scheme's type for \scheme{filter},
type-checking this expression is now straightforward. \scheme|filter|
can of course be user-defined, the straightforward implementation is
accepted with the above type.
The example again demonstrates
type inference for local non-recursive bindings.

The type of \scheme|call/cc| must reflect the fact that invoking a
continuation aborts the local computation in progress:

\begin{schemedisplay}
call/cc : (All (a) (((a -> bot) -> a) -> a))
\end{schemedisplay}

\noindent
where \scheme|bot| is the empty type, expressing the fact that the function
cannot produce values. This type has the same logical interpretation as
Peirce's law, the conventional type for
\scheme|call/cc|~\cite{griffin:popl91} but works better with our type inference system.

\begin{figure}[t!]
\begin{schemedisplay}
lang typed-scheme
(provide LoN sum)
(define-type-alias LoN (Listof Number))
(define: (sum [l : LoN]) : Number
  (if (null? l) 0 (+ (car l) (sum (cdr l))))))
\end{schemedisplay}

\hrule

\begin{schemedisplay}
lang typed-scheme
(require m1)
(define: l : LoN (list 1 2 3 4 5))
(display (sum l)))
\end{schemedisplay}
\caption{A Multi-Module Typed Scheme Program}
\label{f:multi}
\end{figure}

\subsection{Programming in the Large}

PLT Scheme
has a first-order module system~\cite{f:modules} that
allows us to support multi-module typed programs with no extra effort.
In untyped PLT Scheme programs, a module consists of definitions and
expressions, along with declarations of dependencies on other modules, and
of export specifications for identifiers.  In Typed Scheme, the same module
system is available, without changes.  Both defined values and types can be
imported or provided from other Typed Scheme modules, with no syntactic
overhead.  The types of provided identifiers is taken from their
initial definition.
In the example in figure \ref{f:multi}, the type \scheme|LoN| and the function \scheme|sum|
are provided by module \scheme|m1| and  can therefore be used in
module \scheme|m2|
at their declared types.

Additionally, a Typed Scheme module, like a PLT Scheme module, may contain
and export macro definitions that refer to identifiers or types defined in
the typed module.





\subsection{Interoperating with Untyped Code}

\textbf{Importing from the Untyped World}
When a typed module must import functions from an untyped module---say PLT
Scheme's extensive standard library---Typed Scheme requires dynamic checks
at the module boundary. Those checks are the means to enforce type
soundness~\cite{thf:dls2006}.  In order to determine the correct
checks and in keeping with our decision that only binding positions in
typed modules come with type annotations, we have designed a typed import
facility. For example, 
\begin{schemedisplay}
(require/typed scheme [add1 (Number -> Number)])
\end{schemedisplay}
imports the \scheme|add1| function from the \scheme|scheme| library, with
the given type. The \scheme{require/typed} facility expands into contracts,
which are enforced as values cross module
boundaries~\cite{ff:ho-contracts}.  In this example, the use of
\scheme|require/typed| is automatically 
rewritten to a plain \scheme|require| along
with a contract application using the contract
 \scheme|(number? . -> . number?)|.


An additional complication arises when an untyped module provides an opaque
data structure, i.e., when a module exports constructors and operators on
data without exporting the structure definition. In these cases, we do not
wish to expose the structure merely for the purposes of type
checking. Still, we must have a way to dynamically check this type at the
boundary between the typed and the untyped code and to check the typed
module. 

For these situations, Typed Scheme supports  {\it opaque types},
 in which only the predicate for testing membership is specified.
This predicate can be trivially turned into a contract, but no operations
on the type are allowed, other than those imported with the appropriate
type from the untyped portion of the program.  Of course, the predicate is
naturally integrated into the occurrence type system, allowing modules to
discriminate precisely the elements of the opaque type.

Here is a sample usage of the special form for importing a predicate and
thus defining an opaque type:
\begin{schemedisplay}
(require/typed [opaque xml Doc document?])
\end{schemedisplay}
It imports the \scheme|document?| function from the \scheme|xml|
 library and uses it to define the \scheme|Doc| type. The rest of the module
can now import functions with \scheme{require/typed} that refer to \scheme{Doc}.

\textbf{Exporting to the Untyped World}
When a typed module is required by untyped code, the typed code must
be protected \cite{thf:dls2006}. Since exports from typed code come
equipped with a type, they are 
 automatically guarded by contracts, without
additional effort or annotation by the programmer.  Unfortunately,
because macros allow unchecked access to the internals of a module,
macros defined in a typed module cannot currently be imported into an untyped context.

\section{Implementation}
\label{s:impl}

We have implemented Typed Scheme as a language for the PLT Scheme
environment, and it is available in the standard PLT Scheme distribution~\cite{ctf:macros}.
\footnote{The implementation consists of approximately 10000 lines of
  code and 6800 lines of tests.}  The implementation is available from
\url{http://www.plt-scheme.org} .

Since Typed Scheme is intended for use by programmers developing real
applications, a toy implementation was not an option.  Fortunately, we
were able to implement all of Typed Scheme as a layer on top of PLT
Scheme, giving us a full-featured language and standard library.  In
order to integrate with PLT Scheme, all of Typed Scheme is implemented
using the PLT Scheme macro system~\cite{ctf:macros}. When the macro
expander finishes successfully, the program has been typechecked, and
all traces of Typed Scheme have been compiled away, leaving only
executable PLT Scheme code remaining.  The module can then be run
just as any other Scheme program, or linked with existing modules.

\subsection{Changing the Language}

Our chosen implementation strategy requires an integration of the type
checking and macro expansion processes.

The PLT Scheme macro system allows language designers to
control the macro expansion process from the top-most abstract syntax node.  Every PLT Scheme module
takes the following form:

\begin{schemedisplay}
(module m language
   ...)
\end{schemedisplay}
\noindent
where \scheme|language| can specify any library.  The library is then
used to provide all of the core Scheme forms.  For our purposes, the
key form is \scheme|#
which is wrapped around the entire contents of the module, and
expanded before any other expansion or evaluation occurs.  Redefining
this form gives us complete control over the expansion of a Typed
Scheme program. At this point, we can typecheck the module and signal
an error at macro-expansion time if it is ill-typed.

\subsection{Handling Macros}

One consequence of PLT Scheme's powerful macro system is that a large
number of constructs that might be part of the core language are
instead implemented as macros.  This includes pattern
matching~\cite{match}, class systems~\cite{fff:classes} and component systems~\cite{ff:units}, as well
as numerous varieties of conditionals and even boolean operations such
as \scheme|and|.  Faced with this bewildering array of syntactic
forms, we could not hope to add each one to our type system,
especially since new ones can be added by programmers in libraries or
application code.  Further, we cannot abandon macros---they are used
in virtually every PLT Scheme program, and we do not want to require
such changes.  Instead, we transform them into simpler code.

In support of such situations, the PLT Scheme macro system provides the
\scheme|local-expand| primitive, which expands a form in the current
syntactic environment.  This allows us to fully expand the original
program in our macro implementation of Typed Scheme, prior to type
checking.  We are then left with only the PLT Scheme core forms, of
which there are approximately a dozen.








\subsection{Cross-Module Typing}

In PLT Scheme programs are divided up into first-order modules.  Each
module explicitly specifies the other modules it imports and
the bindings it exports.  In order for Typed Scheme to work
with actual PLT Scheme programs, it must be possible for
programmers to split up their Typed Scheme programs into multiple
modules.  

Our type-checking strategy requires that all
type-checking take place during the expansion of a particular module.
Therefore, the type environment constructed during the typechecking of
one module disappears before any other module is considered.  

Instead, we turn the type environments into persistent code using Flatt's
reification strategy~\cite{f:modules}.  After typechecking each module, the type
environment is reified in the code of the module as instructions for
recreating that type environment when that module is expanded.
 Since every dependency of a module is visited during the expansion of
 that module, the appropriate type environment is recreated for each
 module that is typechecked.  
This implementation technique has the significant benefit that it
provides separate compilation and typechecking of modules for free.

 Further, our type environments are keyed by PLT Scheme identifiers,
 which maintain information on which module they were defined in,
 providing several advantages.
First, the technique described by \citet{f:modules} and adapted
 for Typed Scheme by \citet{ctf:macros} allows the use
 of one typed module from
 another without having to redeclare types. Second, standard tools
 for operating on PLT Scheme programs,
 such as those provided by DrScheme~\cite{fffkf:drscheme-journal} work
 properly with typed programs and binding of types.

\subsection{Performance}

There are three important aspects to the performance of Typed Scheme:
the performance of the typechecker itself, the overhead of contracts
generated for interoperation, and the overhead that Typed Scheme's
runtime support imposes on purely typed program.  We address each in
turn.  

The typechecker is currently notably slower than macro expansion
without typechecking, but is not problematically slow.  Even large
files typecheck in just a few seconds.  We have optimized the
typechecker significantly over the development of Typed Scheme; the
most significant optimization is interning of all type
representations, allowing constant-time type comparison and
substantially reducing memory use.  

The overhead of contracts can be substantial, depending on the
particular contracts generated.  In some cases, contracts can change
the asymptotic complexity of existing programs.  We hope to investigate
techniques for lazy checking of
contracts~\cite{findler:contracts-lazy} to alleviate this problem.
However, this overhead is only imposed when crossing the typed-untyped
boundary, which we predict will be rare in inner loops and other
performance critical code.  Adding types to selected portions of the
DrScheme~\cite{fffkf:drscheme-journal} implementation resulted in no
measurable slowdown.

Finally, the implementation of Typed Scheme imposes no runtime overhead
on programs, with the exception of the need to load the code
associated with the library.  Thus typed code executes at full speed.
We are investigating optimization opportunities based on the type
information~\cite{stamour:opt}.  

\subsection{Limitations}

Our implementation has two significant limitations at present.  First,
we are unable to dynamically enforce some types using the PLT contract
system.  For example, although checking polymorphic
types~\cite{gmfk07,afmw:forall} is supported, variable-arity
polymorphism is not.  Additionally, mutable data continues to present
problems for contracts.  As solutions for these limitations are
integrated into the PLT Scheme contract system, more of Typed Scheme's
types will be dynamically enforceable.

The second major limitation is that we cannot typecheck code that uses
the most complex PLT Scheme macros, such as the \scheme|unit| and
\scheme|class| systems.  
These macros maintain their own invariants, which must be understood
by the typechecker in order to sensibly type the program.  For
example, the \scheme|class| macro maintains a \emph{vtable} for each
class, which in this implementation is a set of methods indexed by
symbols.  Typing such an implementation would require either a
significant increase in the complexity of the type system or special
handling of such macros. 
Since these
macros are widely used by PLT Scheme programmers, we plan to
investigate both possibilities.

\section{Practical Experience}
\label{s:exp}

To determine whether Typed Scheme is practical and whether converting
PLT Scheme programs is feasible, we conducted a series of experiments
in porting existing Scheme programs of varying complexity to Typed
Scheme.

{\bf Educational Code}
For small programs, which we expected to be written in a disciplined
style that would be easy to type-check, we turned to educational
code.  Our preliminary investigations and type system design indicated
that programs in the style of {\it How to Design Programs}~\cite{fffk:htdp}
would type-check successfully with our system, with only type
annotations required.  

To see how more traditional educational Scheme code would fare, we
rewrote most programs from  two additional text books: {\it The Little
  Schemer\/}~\cite{little-schemer} and {\it The Seasoned
  Schemer\/}~\cite{seasoned-schemer}.  Converting
these 500 lines of code usually required nothing but the declaration
of types for function headers.  The only difficulty encountered was
 an inability to express in our type system some
invariants on S-expressions that the code relied on.

Second, we ported 1,000 lines of educational code, which consisted of
the solutions to a number of exercises for an undergraduate
programming languages course.  Again, handing S-expressions proved the
greatest challenge, since the code used tests of the form
\scheme|(pair? (car x))|, which does not provide useful information to
the type system (formally, the visible predicate of this expression is
\noeffect).  Typing such tests required adding new local bindings.
This code also made use of a non-standard datatype definition
facility, which required adaptation to work with Typed Scheme.

{\bf Libraries}
We ported 500 lines of code implementing a variety of data
structures from S\o gaard's~\nocite{galore}
\scheme|galore.plt| library package. While these data structures were
originally designed for a typed functional language, the
implementations were not written with typing in mind. Two sorts of
changes were required for typing this library. First, in several
places the library failed to check for erroneous input, resulting in potentially surprising
behavior.  Correcting this required adding tests for the erroneous
cases.  Second, in about a dozen places throughout the code,
polymorphic functions needed to be explicitly instantiated in order
for typechecking to proceed.  These changes were, again, in addition
to the annotation of bound variables.

{\bf Applications}
A research intern ported two
sizable applications under the direction of the first author.  The
first was a 2,700 line implementation of a game, written in 2007, and
the second was a 500 line checkbook managing script, maintained for
 12 years.

The game is a version of the multi-player card game
Squadron Scramble.\footnote{Squadron Scramble resembles Rummy; it is
available from US Game Systems.}  The original implementation consists of
10 PLT Scheme modules, totaling 2,700 lines of implementation code,
including 500 lines of unit tests.  

A representative function definition from the game is given in figure
\ref{f:squad}.  This function creates a \scheme|turn| object, and hands it
to the appropriate \scheme|player|.  It then checks whether the game is
over and if necessary, constructs the new state of the game and returns it.

The changes to this complex function are confined to
the function header.  We have converted the original \scheme|define|
to \scheme|define:| and provided type annotations for each of the
formal parameters as well as the return type.  This function 
returns multiple values, as is indicated by the return type.  Other
than the header, no changes are required.  The types of all the
locally bound variables are inferred from the bodies of the individual
definitions.  

Structure types are used extensively in this example, as well as in
the entire implementation.  In the definition of the variables
\scheme|the-end| and \scheme|the-return-card|, occurrence typing is
used to distinguish between the \scheme|res| and \scheme|end|
structures.   

Some portions of the implementation required more effort to port to
Typed Scheme. For example, portions of the data used for the game is stored
in external XML files with a fixed format, and the program relies upon
the details of that format.  However, since this invariant is neither
checked nor specified in the program, the type system cannot verify it.  Therefore,
we moved the code handling the XML file into a separate, untyped module of fewer than 50 lines
that the typed portion uses via \scheme|require/typed|.

\begin{figure}
\schemeinput{squad.scm}
\caption{A Excerpt from the Squadron Scramble Game}
\label{f:squad}
\end{figure}

{\bf Scripts}
The second application ported required similarly few changes.  This
script maintained financial records recorded in an S-expression
stored in a file.  The major change made to the program was the
addition of checks to ensure that data read from the file was in the
correct format before using it to create the relevant internal
data structures.  This was similar to the issue encountered with the
Squadron Scramble game, but since the problem concerned a single
function, we added the necessary checks rather than creating a new
module.  The other semantic change to the program was to maintain a
typing invariant of a data structure by construction, rather
than after-the-fact mutation.  As in the case of the Galore library,
we consider this typechecker-mandated change an improvement to
the original program, even though it has already been used successfully
for many years.


\section{Related Work}

The history of programming languages knows many attempts to add or to use
 type information in conjunction with untyped languages. Starting with
 LISP~\cite{cl}, language designers have tried to include type
 declarations in such languages, often to help compilers, sometimes to
 assist programmers. From the late 1980s until recently, people have
 studied soft typing~\cite{cf:pldi91,awl:popl94,wc:toplas97,hr:fpca95,ff:toplas99,mff:popl06}, a
 form of type inference to assist programmers debug their programs
 statically.  This work has mainly been in the context of Scheme but has also been applied to
 Python~\cite{some-mit-ms-thesis}. Recently, the slogan of ``gradual
 typing'' has resurrected the LISP-style annotation mechanisms and has had
 a first impact with its tentative inclusion in Perl6~\cite{tang}. 

In this section, we survey this body of work, starting with the
soft-typing strand, because it is the closest relative of Typed
Scheme. We conclude with a discussion of refinement types.

\subsection{Types for Scheme}

The goal of the soft typing research agenda is to provide an optional type
 checker for programs in untyped languages. One key premise is that
 programmers shouldn't have to write down type definitions or type
 declarations. Soft typing should work via type inference only, just like
 ML. Another premise is that soft type systems should never prevent
 programmers from running any program. If the type
 checker encounters an ill-typed program, it should insert run-time checks that
 restore typability and ensure that the type system remains
 sound. Naturally, a soft type system should minimize these insertions of
 run-time checks. Furthermore, since these insertions represent potential
 failures of type invariants, a good soft type system must allow programmer
 to inspect the sites of these run-time checks to determine whether they
 represent genuine errors or weaknesses of the type system. 

Based on the experiences of the second author, soft type systems are  complex
 and brittle. On one hand, these systems may infer extremely large types
 for seemingly simple expressions, greatly confusing the original
 programmer or the programmer who has taken on old code. On the other hand, a
 small syntactic change to a program without semantic consequences can
 introduce vast changes into the types of both nearby and remote
 expressions. Experiments with undergraduates---representative of 
 average programmers---suggest that only the very best understood the tools
 well enough to make sense of the inferred types and to exploit them for
 the assigned tasks. For the others, these tools turned into time sinks with
 little benefit.

Roughly speaking, soft typing systems fall into one of two classes, depending
 on the kind of underlying inference system. The first soft type
 systems~\cite{cf:pldi91,wc:toplas97,hr:fpca95, h:dynamic} used inference engines
 based on Hindley-Milner though with extensible record types. These systems
 are able to type many actual Scheme programs, including those using
 outlandish-looking recursive datatypes. Unfortunately, these systems
 severely suffer from the general Hindley-Milner error-recovery
 problem. That is, when the type system signals a type error, it is
 extremely difficult---often impossible---to decipher its meaning and to
 fix it.


In response to this error-recovery problem, others built inference systems
 based on Shiver's control-flow analyses~\citeyearpar{shivers-thesis} and
 Aiken's and Heintze's set-based analyses~\cite{awl:popl94,nh:sba}. Roughly
 speaking, these soft typing systems introduce sets-of-values constraints
 for atomic expressions and propagate them via a generalized
 transitive-closure propagation~\cite{awl:popl94,ff:toplas99}. In this
 world, it is easy to communicate to a programmer how a values might flow
 into a particular operation and violate a type invariant, thus eliminating
 one of the major problems of Hindley-Milner based soft
 typing~\cite{ffkwf:mrspidey}.

Our experience and evaluation suggest that Typed Scheme works
 well compared to soft typing. First, programmers can easily
 convert entire modules with just a few type declarations and annotations
 to function headers. Second, assigning explicit types and rejecting programs
 actually pinpoints errors better than soft typing systems, where
 programmers must always keep in mind that the type inference system is
 conservative. Third, soft typing systems do not support type
 abstractions well. Starting from an explicit,
 static type system for an untyped language should help introduce these
 features and deploy them as needed. 

The Rice University soft typing research inspired occurrence typing.  These
 systems employed \scheme{if}-splitting rules that performed a case
 analysis for types based on the syntactic predicates in the test
 expression. This idea was derived from \citet{c:typed-lisp}'s \texttt{typecase} construct (also see below) and---due to its
 usefulness---is generalized by our framework. The major advantage of soft
 typing over an explicitly typed Scheme is that it does not require any
 assistance from the programmer. In the future, we expect to borrow
 techniques from soft typing for automating some of the conversion process
 from untyped modules to typed modules.

 \citet{shivers-thesis} presented 0CFA, which also uses flow analysis
 for Scheme programs.  He describes a possible extension to account
 for occurrence-typing like behavior for literal applications of the
 predicate \scheme|number?|, but did not discuss more general aspects
 of the issue.

\citet{hr:fpca95} used a flow
analysis to convert Scheme programs to ML programs, while minimizing
runtime checks.  While this is also converting Scheme programs to
typed programs, it is intended as a compilation step, not a
refactoring, and the ML code is not intended to be maintained as the
primary form of the program.  Additionally, their system does not
take predicate tests into account, which is the primary focus
of occurrence typing.

\citet{awl:popl94} describe a type inference system using
\emph{conditional types}, which refine the types of variables based on
patterns in a {\ttfamily case} expression.  Since this system is built on
the use of patterns, abstracting over tests as the {\sc T-AbsPred}
rule does, or combining them, as with \scheme|or| is impossible.

\subsection{Gradual Typing}

Under the name ``gradual typing'', several other researchers  have
experimented with the integration of typed and untyped
code~\cite{st:gradual06, Herman07, wadler-findler, stop09}.  This work
has been pursued in two directions.  First, theoretical investigations
have considered integration of typed and untyped code at 
 a much finer granularity than we
present, providing soundness theorems which prove that only the
untyped portions of the program can go wrong.  This is analogous to our earlier work on Typed
Scheme~\cite{thf:dls2006}, which provides such a soundness theorem, which we
believe scales to full Typed Scheme and PLT Scheme.  These gradual
typing systems have not been scaled to full implementations.

Second, \citet{furr:ruby:sac, furr:ruby:stop} have implemented a
system for Ruby which is similar to Typed Scheme.  They have also
designed a type system which matches the idioms of the underlying
language, and insert dynamic checks at the borders between typed and
untyped code.  Their work does not yet have a published soundness
theorem, and requires the use of a new Ruby interpreter, whereas Typed
Scheme runs purely as a library for PLT Scheme.

\citet{bracha:pluggable} suggests pluggable typing systems, in which a
programmer can choose from a variety of type systems for each piece of
code.  Although Typed Scheme requires some annotation, it can be thought of
as a step toward such a pluggable system, in which programmers can choose
between the standard PLT Scheme type system and Typed Scheme on a
module-by-module basis.

\subsection{Type System Features}

Many of the type system features we have incorporated into Typed
Scheme have been extensively studied.  Polymorphism in type systems
dates to \citet{reynolds}.  Recursive types are considered by
\citet{ac:subtyping-recursive-types}, and union
types by \citet{pierce-union}, among many others. Intensional
polymorphism appears in calculi by \citet{harper95compiling}, among others. Our use of
\lexeff{}s and especially \lateff{}s is inspired by prior work on
effect systems~\cite{fx87}.  

\subsection{Other Type Systems}

 \citet{c:typed-lisp} describes Typed Lisp, which includes
{\texttt{typecase}} expression that refines the type of a variable in the
various cases; \citet{weirich98} re-invent this construct in the context of
a typed lambda calculus with intensional polymorphism.  The
{\texttt{typecase}} statement specifies the variable to be refined, and
that variable is typed differently on the right-hand sides of the
\texttt{typecase} expression.  While this system is superficially similar
to our type system, the use of latent and visible predicates allows us
to handle cases other than simple uses of \texttt{typecase}.  This is important
in type-checking existing Scheme code, which is not written with
\texttt{typecase} constructs. 

Visible predicates can also be seen as a kind of dependent type, in
that \scheme|(number? e)| could be thought of as having type \tt when
\scheme|e| has a value that is a number.  In a system with singleton
types, this relationship could be expressed as a dependent type.
This kind of combination typing would not cover the use of \scheme|if| to refine
the types of variables in the branches, however.  

The term ``occurrence typing'' was coined independently by
\citet{krcf:occurrence}, in the context of a static analysis system
for Cobol.  That system considers a specific syntactic form of
\scheme|if| tests: the comparison of variables with character
literals.  This accommodates a common encoding of datatypes in Cobol
programs.  It does not allow for abstraction over tests or any other
form of predicates.

\subsection{Type Systems for Untyped Languages}

Multiple previous efforts have attempted to typecheck Scheme programs.
\citet{wand84}, \citet{h:infer}, and \citet{leavens}
developed typecheckers for an ML-style type system, each of which
handle polymorphism, structure definition and a number of Scheme
features.  Wand's macro-based system integrated with untyped Scheme code via
unchecked assertions.  Haynes' system also handles variable-arity
functions~\cite{var-ar}.  However, none attempts to accommodate a
traditional Scheme programming style.

Bracha and Griswold's Strongtalk~\citeyearpar{strongtalk}, like Typed Scheme, presents
a type system designed for the needs of an untyped language, in their
case Smalltalk. Reflecting the differing underlying languages, the
Strongtalk type system differs radically from ours and does not
describe a mechanism for integrating with untyped code.

\subsection{Refinement Types}

Refinement types were originally introduced by \citet{fp:refinement}.
Since then, refinement types have been used in a wide variety of
systems~\cite{liquid-types,wadler-findler,f:hybrid}.  Previous
refinement type systems come in two varieties.  Freeman and Pfenning's
original system used the underlying language of ML types to specify
subsets of the existing types, such as non-empty lists, defined by
recursive datatype-like specifications.  Most other
systems have paired predicates in some potentially-restricted language
with a base type, meaning the set of values of that base type accepted
by that predicate.  Typically, this requires some algorithm for
deciding implication between predicates for subtyping.  In some
languages, this can be an external and almost always incomplete theorem
prover, as in the Liquid Typing and Hybrid Typing approaches.  

Typed Scheme provides support for both of these approaches, as seen in
section~\ref{sec:example}.  To support Freeman and Pfenning's style,
such data types can be directly encoded via recursive types.  Typed
Scheme is able to handle all of Freeman and Pfenning's examples in
this fashion.
To support predicate style-refinement, Typed Scheme takes a different
approach.
  First, refinements are not specified using a special language
of  predicates or formulae but as in-language predicates.  This
allows any computable set to be a refinement.  Second,
no attempt is made to decide implication between predicates.  Two
distinct functions might be extensionally equivalent, but the
associated refinement types have no subtyping relationship.  This
frees both the programmer and the implementor from the burden of
depending on a theorem prover.


\section{Conclusion}

Migrating programs from untyped languages to typed languages is an important
problem. In this paper we have demonstrated one successful approach, based
on the development of a type system that accommodates the idioms and
programming styles of our scripting language of choice.

Our type system combines a simple new idea, occurrence typing, with a range
of previously studied type system features with some widely used  and some only
studied in theory. Occurrence typing assigns distinct subtypes of a
parameter to distinct occurrences, depending on the control flow of the
program. We introduced occurrence typing because our past experience
suggests that Scheme programmers combine flow-oriented reasoning with
typed-based reasoning.  Occurrence typing also allows us to naturally
extend the type system with a simple and expressive form of refinement
types, allowing for static verification of arbitrary property checking.

Building upon this design, we have implemented and distributed Typed
Scheme as a package for the PLT Scheme system.  This implementation
supports the key type system features discussed here, as well as
integration features necessary for interoperation with the rest of the PLT
Scheme system.  

Using Typed Scheme, we have evaluated our type system. We consider the
experiments of section \ref{s:exp} illustrative of existing code and
believe that their success is a good predictor for future experiments.
We plan on continuing to port PLT Scheme libraries to Typed Scheme and on
exploring the theory of occurrence typing in more depth.

For a close look at Typed Scheme, including documentation and sources for
its Isabelle/HOL and PLT Redex models, visit the Typed Scheme
web page:

\begin{center}
\verb|http://www.ccs.neu.edu/~samth/typed-scheme|
\end{center}

\end{schemeregion}

\begin{acknowledgements}
We thank Ryan Culpepper for invaluable assistance with the
implementation of Typed Scheme, Matthew Flatt for implementation
advice, Ivan Gazeau for his porting of existing PLT Scheme code, and
members of Northeastern PRL as well as
several anonymous  reviewers for their comments.
\end{acknowledgements}

\bibliographystyle{spbasic}      
\bibliography{new-popl,extra}   

\end{document}